%% file: WW_PRD.tex
\documentclass[aps,prd,twocolumn,showpacs,superscriptaddress,groupedaddress]{revtex4}  

\usepackage{graphicx}
\usepackage{dcolumn}
\usepackage{bm}
\usepackage{lineno}
\usepackage{setspace}
\begin{document}



\vspace*{1.5cm}

\title{Measurement of the production and differential cross sections of $W^{+}W^{-}$ bosons in association with jets in $p\bar{p}$ collisions at $\sqrt{s}=1.96$ TeV}
\input{author.tex}

\vspace*{2.0cm}

\begin{abstract}
We present a measurement of the $W$-boson-pair production cross section in $p\bar{p}$ collisions at 1.96 TeV center-of-mass energy and the first measurement of the differential cross section as a function of jet multiplicity and leading-jet energy.  The $W^{+}W^{-}$ cross section is measured in the final state comprising two charged leptons and neutrinos, where either charged lepton can be an electron or a muon.  Using data collected by the CDF experiment corresponding to $9.7~\rm{fb}^{-1}$ of integrated luminosity, a total of $3027$ collision events consistent with $W^{+}W^{-}$ production are observed with an estimated background contribution of $1790\pm190$ events.  The measured total cross section is $\sigma(p\bar{p} \rightarrow W^{+}W^{-}) = 14.0 \pm 0.6~(\rm{stat})^{+1.2}_{-1.0}~(\rm{syst})\pm0.8~(\rm{lumi})$ pb, consistent with the standard model prediction.
\end{abstract}

\maketitle

\newpage
The measurement of $W$-boson-pair production is an important test of the electroweak gauge sector of the standard model (SM) of particle physics~\cite{Ellison:1998uy,Hagiwara:1989mx}.  Extending the measurement to include the properties of jets produced in association with the boson pair
is an interesting test of quantum chromodynamics (QCD) that has not been performed before in events with multiple heavy gauge bosons.  This process is a significant background for Higgs-boson studies in the $W^{+}W^{-} (WW)$ decay mode at particle colliders, where the use of jet vetoes and jet counting is an essential element of the analysis technique~\cite{Aaltonen:2013iia,Abazov:2013wha,Aad:2012uub,Chatrchyan:2013iaa}. 
Furthermore, a final state consisting of two massive gauge bosons and two jets is typical of vector-boson scattering, a process sensitive to many non-SM contributions to the electroweak-symmetry-breaking sector~\cite{Englert:2008tn}.  
With the first evidence of vector-boson scattering at the Large Hadron Collider (LHC)~\cite{Aad:2014zda,Khachatryan:2014sta}, it is essential to verify and tune the simulation tools used to describe electroweak processes produced in association with jets.\\  
\indent This communication reports measurements of the $W^{+}W^{-}$ production and differential cross sections as a function of jet multiplicity and jet energy in a final state consisting of events with two oppositely charged leptons, where each lepton is identified as either an electron or a muon, and an imbalance in the total event transverse momentum (transverse energy), due to the presence of neutrinos.   The differential measurements of jet multiplicity and jet energy are unfolded to hadronic-jet level
to account for the detector response to jets of hadrons and compared directly to predictions from two Monte Carlo (MC) simulations.  The simulations considered are the {\sc alpgen} generator~\cite{Mangano:2002ea}, 
which calculates the lowest perturbative order of the strong interaction that produces a $W^{+}W^{-}$ pair of vector bosons and a fixed number of partons in the final state  (typically referred to as fixed-order MC), and the {\sc mc@nlo} next-to-leading order (NLO) generator~\cite{Frixione:2002ik} both interfaced to parton-shower generators.  Fixed-order and NLO simulation are widely used methods for modeling electroweak processes with multiple jets.  The measurement of
differential cross sections as a function of the number of associated jets is particularly suited to the Tevatron because the large rate of top-quark-pair production at the LHC makes this analysis much more difficult.
The production of $W$-boson pairs was first observed by the CDF experiment using Tevatron Run I data~\cite{Abe:1996dw}. 
This process has since been measured by the CDF~\cite{Aaltonen:2009aa}, D0 ~\cite{Abazov:2009ys}, ATLAS~\cite{Aad:2011kk}, and CMS~\cite{Chatrchyan:2013yaa,Chatrchyan:2013oev} experiments.\\ 
\indent  This measurement uses an integrated luminosity of $9.7~\rm{fb}^{-1}$ of $p\bar{p}$ collision data collected by the CDF experiment which comprises the full Tevatron Run II data set.
A 12\% increase in signal acceptance is obtained by using events in which one or both $W$ bosons decay to a tau lepton that subsequently decays to an electron or muon.
Events are classified based on the multiplicity and the transverse energy of jets, where jets are the products of the parton showering and hadronization of high energy partons produced in the initial scattering process.
The cross section measurement uses an artificial neural network (NN) method to distinguish signal and background events.  The jet multiplicity classifications are zero, one, and two or more jets and separate NNs are used for each multiplicity class.  Events with a single jet are further classified according to the energy of the jet transverse to the beam line into bins of $15-25$ GeV, $25-45$ GeV, and more than $45$ GeV. 
The cross sections are extracted via a maximum-likelihood fit to the data of a weighted sum of the normalized binned NN distributions for signal and backgrounds, simultaneously over
the five event classes.\\
 \indent
The CDF experiment consists of a solenoidal spectrometer with a silicon tracker and an open-cell drift chamber surrounded by calorimeters and muon detectors~\cite{Acosta:2004yw}.  The kinematic properties of particles and jets are characterized using the azimuthal angle $\phi$ and the pseudorapidity $\eta \equiv - \ln(\tan(\theta/2))$, where $\theta$ is the polar angle relative to the nominal proton beam axis.  Transverse energy, $E_T$, is defined to be $E\sin\theta$, where $E$ is the energy deposited in pointing-tower geometry electromagnetic and hadronic calorimeters.  Transverse momentum, $p_T$, is the momentum component of a charged particle transverse to the beam-line. \\
 \indent  This analysis uses jets, electrons, muons and missing transverse energy.  Electron and muon candidates (hereafter electrons and muons) are typically identified using the drift chamber and electromagnetic calorimeter or muon chambers, respectively.  
Jet candidates (jets) are measured using an iterative cone algorithm that clusters signals from calorimeter towers within a cone of $\Delta R = \sqrt{(\Delta\eta)^2+(\Delta\phi)^2} = 0.4$. Corrections are applied to the calorimeter cluster-energy sums to account for the average calorimeter response to jets~\cite{Bhatti:2005ai}. The missing transverse energy vector, $\vec{E\kern-0.50em\raise0.10ex\hbox{/}_{T}}$, is defined as the opposite of the vector sum of the $E_T$ of all calorimeter towers, corrected to account for the calorimeter response to jets and to muons\cite{Aaltonen:2013iia}. \\
 \indent
The experimental signature of the signal events is two leptons with opposite charge and large $E\kern-0.50em\raise0.10ex\hbox{/}_{T}$.  A number of SM processes result in this observed final state and are thus backgrounds.  The {\it WZ} and {\it ZZ} processes can yield the same final state if one boson decays hadronically, or some decay products are not detected. Top quarks decay predominantly to a $b$ quark and a $W$ boson, which can decay leptonically, making $t\bar{t}$ production a dominant background for events with two jets.  
Incorrect measurements of the energy or momentum of leptons or jets result in apparent $E\kern-0.50em\raise0.10ex\hbox{/}_{T}$.  
This allows the large-production-rate Drell-Yan (DY) process, $Z/\gamma^{*} \rightarrow \ell^{+}\ell^{-}$, where $\ell$ refers to any charged lepton, to be included in the candidate sample.
A third source of background arises from the misidentification of a final-state particle, such as in $W+$jets and $W\gamma$ processes, where the photon or jet is incorrectly reconstructed as a lepton. \\
 \indent  Events are selected from a sample containing at least one electron (muon) of $E_T(p_T) > 20$ GeV  and $|\eta|< 2.0$ that satisfy an online single-lepton selection requirement in the trigger~\cite{Acosta:2004uq}.  
A second lepton is required to have opposite charge and $E_T (p_T) > 10$ GeV.  
Jets are required to have $E_T > 15$ GeV and $|\eta| < 2.5$.  
In order to reduce the contribution of DY events with mismeasured $E_T (p_T)$, the $E\kern-0.50em\raise0.10ex\hbox{/}_{T,{\it rel}}$ is defined as $E\kern-0.50em\raise0.10ex\hbox{/}_{T,{\it rel}} \equiv E\kern-0.50em\raise0.10ex\hbox{/}_{T}\sin\Delta\phi(\vec{E\kern-0.50em\raise0.10ex\hbox{/}_{T}},\ell/j)$ if the azimuthal separation between the $\vec{E\kern-0.50em\raise0.10ex\hbox{/}}_{T}$ and the momentum vector of the nearest lepton or jet is less than $\frac{\pi}{2}$.  Otherwise $E\kern-0.50em\raise0.10ex\hbox{/}_{T,{\it rel}} = E\kern-0.50em\raise0.10ex\hbox{/}_{T}$.  $E\kern-0.50em\raise0.10ex\hbox{/}_{T,{\it rel}}$ is required to be greater than $25$ GeV for same-flavor lepton pairs, or $15$ GeV for electron-muon events from the leptonic decays of $\tau^+\tau-$ DY events, where the DY contribution is small.  
Requiring the invariant mass of the same-flavor lepton pairs $(m_{\ell\ell})$ to be outside the $Z$ mass range (between $80$ and $99$ GeV/$c^2$) further suppresses
$Z/\gamma^{*} \rightarrow \ell^{+}\ell^{-}$ events, while requiring one radian of azimuthal separation between the $E\kern-0.50em\raise0.10ex\hbox{/}_{T}$ and the dilepton momentum suppresses $Z/\gamma^{*} \rightarrow \tau^{+}\tau^{-}$ events.  Leptons are required to be isolated, by applying the criteria that the sum of $E_T$ for calorimeter towers (or the sum of the $p_T$ of charged particles for central muons) in a cone of $\Delta R$ around the lepton is less than $10\%$ of the electron $E_T$ (muon $p_T$).  This helps to purify the lepton sample and reduce backgrounds due to misidentified particles, particularly $W+$jets and $W\gamma$.  The $W\gamma$ background is further reduced by requiring $m_{\ell\ell}$ greater than 16 GeV/$c^2$. To suppress $t\bar{t}$ background we reject events with two or more jets in which any jet passes a $b$-flavored-quark identification algorithm~\cite{Freeman:2012uf}.  \\
 \indent
  The geometric and kinematic acceptance for the {\it WW} signal and the yields of the DY, {\it WZ}, {\it ZZ}, $W\gamma$ and $t\bar{t}$ backgrounds is determined by simulation.   Backgrounds consisting of DY leptons associated with the production of zero or one jet are simulated using the {\sc pythia} generator~\cite{Sjostrand:2001yu}.  DY background with two or more jets and {\it WW} samples are simulated with up to three partons with the {\sc alpgen} generator.   The contributions to the {\it WW} production from the next-to-next-to-leading-order (NNLO) process $gg \rightarrow W^+W^-$ are accounted for by reweighting the {\it WW} simulation as a function of the {\it WW} $p_T$ distribution to incorporate the extra contribution predicted in Ref.~\cite{Binoth:2006mf}.  
The DY background is studied in a sample of low $E\kern-0.50em\raise0.10ex\hbox{/}_{T,{\it rel}}$ events and a correction is applied to account for the mismodeling of $\vec{E\kern-0.50em\raise0.10ex\hbox{/}_{T}}$ in the simulation of zero-jet and one-jet events~\cite{Aaltonen:2013iia}.   The {\it WZ}, {\it ZZ}, and $t\bar{t}$ samples are simulated with the {\sc pythia} generator~\cite{Sjostrand:2001yu}.  The expected yields for simulated processes are normalized using cross section calculations performed at NNLO for $t\bar{t}$~\cite{Czakon:2013goa}, NLO for {\it WZ}, {\it ZZ}~\cite{Campbell:1999ah} and DY~\cite{Martin:2003sk}, and fixed order for {\it WW}~\cite{Mangano:2002ea}.  The $W\gamma$ background is simulated using the Baur MC generator~\cite{Baur:1989gk}, but is normalized according to a study of same-charge lepton with low $m_{\ell\ell}$~\cite{Aaltonen:2013iia}.  Samples are generated using the CTEQ5L~\cite{Lai:1999wy} parton distribution functions (PDFs), and interfaced to {\sc pythia} for parton showering, fragmentation, and hadronization.  The detector response for these processes is modeled with a {\sc geant3}-based simulation~\cite{Brun:1987ma}, and further corrected for trigger, reconstruction, and identification efficiencies using samples of $W \rightarrow e\nu$ and $Z \rightarrow \ell\ell$ events~\cite{Acosta:2004uq}. The $W+$jets background is determined using collision data.  The probability for a jet to be reconstructed as a lepton is measured in jet-triggered data samples and applied to a $W+$jets data sample to estimate the contribution from $W+$jets events.  The estimated and observed yields for signal and background in each jet region are shown in Table~\ref{tab:Yields_Fit}. \\
 \indent\begin{table}
\centering
\begin{tabular}{l c c c}
\hline \hline
Process & \multicolumn{3}{c}{Events (best fit)} \\
  & Zero jets & One jet & Two or more jets \\
 \hline
{\it WZ} & $19.5\pm3.0$ & $16.7\pm2.3$ & $4.26\pm0.81$ \\
 {\it ZZ} & $13.2\pm1.9$ & $4.25\pm0.61$ & $1.33\pm0.26$ \\
 $t\bar{t}$ & $3.7\pm1.0$ & $76\pm12$ & $158\pm16$ \\
 $DY$ & $150\pm34$ & $83\pm21$ & $20.2\pm8.6$ \\
 $W\gamma$ & $214\pm27$ & $44.0\pm6.4$ & $7.5\pm1.9$ \\
 $W+$jets & $685\pm118$ & $250\pm46$ & $81\pm15$ \\
 \hline
Total bk & $1086\pm124$ & $474\pm57$ & $272\pm26$ \\
 \hline
{\it WW} & $963\pm108$ & $224\pm29$ & $73\pm20$ \\
 \hline
{\it WW}+bk & $2049\pm177$ & $698\pm73$ & $345\pm39$ \\
 \hline
Data & $2090$ & $682$ & $331$ \\
 \hline \hline
\end{tabular} 
\caption{Estimated and observed event yields.  Event yields and uncertainties are normalized to the values returned by fitting signal and background (bk) distributions to the data. Uncertainties are due to statistical and systematic sources as described in the text. }
\label{tab:Yields_Fit}
\end{table}

\begin{figure}
\begin{center}
\includegraphics[width=0.5\textwidth]{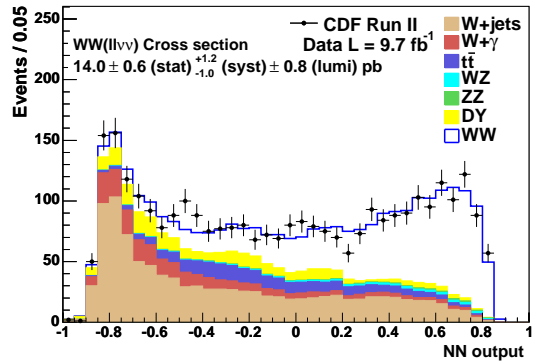}
\caption{Estimated and observed combined distributions of outputs for neutral networks (NNs) trained to identify {\it WW} events, with a higher NN output value indicating a more signal-like event.  Event yields are normalized to the values returned by fitting signal and background distributions to the data.   \label{fig:WW}}
\end{center}
\end{figure}

\indent We use a NN method to further discriminate signal from background.  Separate NNs are trained using the {\sc neurobayes}~\cite{Feindt:2006pm} program with simulated signal and background events for events with zero, one, and two or more jets.  Inputs to the NNs are selected to exploit features of the signal and background processes. The inputs include the total energy of the event, which tends to be modest in {\it WW} events compared to $t\bar{t}$ events; the $p_T$ of the second highest $p_T$ lepton, which tends to be low in events where a jet or photon is misidentified as a lepton; missing $E_T$ that is not aligned with a lepton or jet, suggesting that it is the result of neutrino production; and, in events with two or more jets, the vector sum of the $p_T$ of the leading two jets, which tends to be greater if the jets are $t\bar{t}$ decay products.  For events without jets,
we use a likelihood ratio of the signal probability density over the sum of signal and background probability densities as an input to the NN~\cite{Aaltonen:2013iia}. The probability densities are calculated on an event-by-event basis for the signal and each background hypothesis using the observed kinematic properties of the events in leading-order (LO) matrix-element calculations~\cite{Campbell:1999ah}.  The sum of final NN outputs for all event classes is shown in Fig.~\ref{fig:WW}.\\
 \indent Systematic uncertainties affecting both the normalization and shape of the signal and background NN distributions are assessed for the following sources.  Uncertainties on acceptance originating from lepton-selection and trigger-efficiency measurements contribute a $4.3\%$ uncertainty on all event yields.  Acceptance uncertainties due to potential contributions from higher-order effects are evaluated as follows.  For the {\it WW} contribution, the {\sc alpgen} sample is reweighted by the $p_T$ of the {\it WW} system according to samples generated with different choices of PDF, renormalization, and factorization scales.  The PDF uncertainty is evaluated with the CTEQ61 PDF error set, and ranges from $1.2\%$ to $1.8\%$, increasing as a function of jet multiplicity and jet energy.  The scale uncertainty is evaluated by doubling or halving the renormalization and factorization scales simultaneously and ranges
from $0.5\%$ to $24\%$, increasing as a function of jet multiplicity and decreasing as a function of jet energy.  The scale uncertainties are anticorrelated between 
adjacent bins of jet energy and  jet multiplicity. Between the zero-jet and one-jet bins the anticorrelation is specifically between the bin with no jets and one jet but having the lowest $E_T$.
The effect of the scale uncertainty on the shape of the {\it WW} NN distributions is also evaluated.  For $t\bar{t}$ contribution an uncertainty of $2.7\%$ is assigned due to QCD effects~\cite{Aaltonen:2013tsa}.  For {\it WZ} and {\it ZZ} backgrounds, which are simulated at LO, a $10\%$ uncertainty is assigned, arising from the difference in the observed acceptance of {\it WW} events generated at LO and NLO with {\sc pythia} and {\sc mc@nlo}, respectively.  Because modeling of higher-order amplitudes can affect the extrapolation of the normalization to the predicted $W\gamma$ event yield, this uncertainty is also applied to the $W\gamma$ background.  A $6.8\%$ uncertainty is also assigned to the $W\gamma$ background due to the photon-conversion modeling.  An uncertainty of $19\%$ to $26\%$, increasing as both a function of jet multiplicity and jet energy, is assigned to the DY background to account for mismodeling of $E\kern-0.50em\raise0.10ex\hbox{/}_{T}$ based on the differences in acceptance observed when varying the $E\kern-0.50em\raise0.10ex\hbox{/}_{T}$ template in the study of low $E\kern-0.50em\raise0.10ex\hbox{/}_{T,{\it rel}}$  DY events described above.  Uncertainty on jet modeling for simulated backgrounds varies from $1.0\%$ to $29\%$, and are anticorrelated between jet multiplicity and jet energy and generally increases as a function of jet multiplicity.  For {\it WW} and DY processes, the effect on the shape of the distribution is also evaluated.  For the $W+$jets background, systematic uncertainties of $20\%$ to $30\%$, depending on lepton types, are determined by calculating the misidentification probabilities at different jet-energy thresholds.  Theoretical uncertainties of $6.0\%$ are assigned to {\it WZ} and {\it ZZ} cross sections~\cite{Campbell:1999ah} and $4.3\%$ to the $t\bar{t}$ cross section~\cite{Czakon:2013goa}.  The measured luminosity, which is used to normalize the yield of all signal and background processes modeled using simulations, has an uncertainty of $5.9\%$~\cite{Acosta:2002hx}.  Tables including the systematic uncertainties used in each jet classification and all correlations and anticorrelations are given in Ref.~\cite{Aaltonen:2013iia}.  The sum of uncertainties from statistical and systematic sources is included for each  signal and background for each jet multiplicity in Table~\ref{tab:Yields_Fit}. 
The uncertainties on the background predictions are those after the fit described below.\\
 \indent  The differential {\it WW} cross section is extracted from the NN output shapes incorporating the normalizations and systematic uncertainties of signal and background in each signal region via a binned-maximum-likelihood method.  The likelihood is formed from the Poisson probabilities of observing $n_i$ events in the $i$th bin, in which $\mu_i$ are expected.  Systematic uncertainties are included as normalization parameters on signal and background and subject to constraint by Gaussian terms.  The likelihood is given by

\begin{equation}
\mathcal{L} = \prod_i \frac{\mu_i^{n_i}e^{-\mu_i}}{n_i !}\prod_c e^{\frac{-S_c^2}{2}}
\end{equation}

\noindent where

\begin{equation}
\mu_i = \sum_k \alpha_k \prod_c (1+f_k^c S_c)(N_k)_i .
\end{equation}

\begin{figure}
\begin{center}
\includegraphics[width=0.5\textwidth]{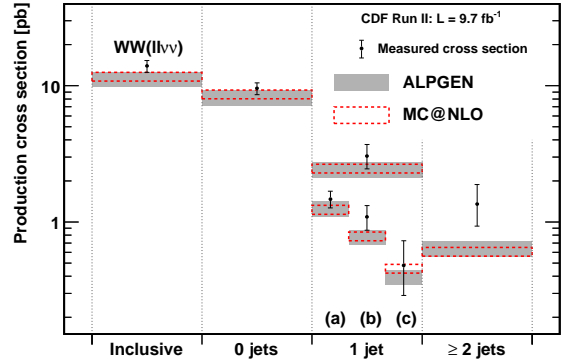}
\caption{Measurement and predictions of $\sigma(p\bar{p} \rightarrow W^{+}W^{-} + \mathrm{n jets})$.  Values are given inclusively and differentially as functions of jet multiplicity and jet-transverse energy.  Transverse energy ranges are (a) $15 < E_T < 25$ GeV, (b) $25 < E_T < 45$ GeV, and (c) $E_T > 45$ GeV. \label{fig:XS}}
\end{center}
\end{figure}

\noindent For a process $k$ and a systematic effect $c$, $f_k^c$ is the fractional uncertainty assigned, and $S_c$ is a Gaussian-constrained floating parameter associated with $c$.  $(N_k)_i$ is the expected number of events from process $k$ in bin $i$.  Systematic uncertainties that affect the shape of the distribution are treated as correlated with the appropriate rate uncertainties.
The parameter $\alpha_k$ is an overall normalization that is fixed to one for all processes except {\it WW}, 
for which it is determined by the fit independently for each analysis region.  The likelihood function is maximized with respect to the systematic parameters $S_c$ and cross section normalizations $\alpha_{WW}$ simultaneously in all regions.  The cross section in each region is calculated by multiplying the value of $\alpha_{WW}$ by the predicted cross sections calculated by {\sc alpgen}.  The total cross section is determined to be $\sigma(p\bar{p} \rightarrow W^{+}W^{-} + X) = 14.0 \pm 0.6\, (\rm{stat})\, ^{+1.2}_{-1.0}\, (\rm{syst})\, \pm0.8\, (\rm{lumi})$ pb which is consistent within one $\sigma$ with the inclusive NLO cross section prediction of $11 \pm 0.7 $ pb as calculated by the {\sc mc@nlo}  program and the total prediction of the fixed-order program {\sc alpgen}.  The result is unfolded to the hadronic-jet level based on the results of a study of the bin-to-bin migration of events due to jet reconstruction, jet-energy scale, and jet-resolution effects as determined in simulated events.  The final result is iteratively corrected to account for differences in acceptance between the reconstructed and true distributions using a Bayesian~\cite{D'Agostini:1994zf,Adye:2011gm} technique.  Migrations between jet-multiplicity and jet-energy bins are typically of order 10\% or less.  An independent training sample of simulated events is used to test the unfolding process. The correct differential cross sections are reproduced stably with a minimal number of iterations.
The unfolded results are compared to {\sc alpgen}, using the CTEQ5L PDFs and interfaced to {\sc pythia} for parton showering with the Michelangelo L. Mangano (MLM) matching algorithm~\cite{Mangano:2006rw}, and {\sc mc@nlo}, using the CTEQ5M PDFs and interfaced to {\sc herwig}~\cite{Corcella:2000bw,Corcella:2002jc} for parton showering. 
The measured and predicted differential cross sections are shown in Table~\ref{tab:XS} and Fig.~\ref{fig:XS}.
The differential cross section measurements are consistent with both simulation predictions.  The largest deviation occurs in the two-jet sample being less that two standard deviations.  NNLO contributions to the 
$q\bar{q} \rightarrow W^+W^-$ cross section are not accounted for in the calculations and are expected to increase the predicted cross section.\\
 \indent
\begin{table*}
\centering
\begin{tabular}{l c c c c c c }
\hline
\hline
   & $\sigma$(pb) & \multicolumn{3}{c}{Uncertainty(pb)} & \multicolumn{2}{c}{$\sigma$(pb)} \\
 Jet bin & Measured & Statistical & Systematic & Luminosity & {\sc alpgen} & {\sc mc@nlo} \\
 \hline
\rule{0pt}{3ex} Inclusive & $14.0$ & $\pm\,0.6$ & $ ^{+1.2}_{-1.0}$ & $\pm\,0.8$ & $11.3\,\pm\,1.4$ & $11.7\,\pm\,0.9$ \\ 
\hline
\rule{0pt}{3ex} Zero jets & $9.57$ & $\pm\,0.40$ & $ ^{+0.82}_{-0.68}$ & $\pm\,0.56$ & $8.2\,\pm\,1.0$ & $8.6\,\pm\,0.6$ \\ 
\hline
\rule{0pt}{3ex} One jet inclusive & $3.04$ & $\pm\,0.46$ & $ ^{+0.48}_{-0.32}$ & $\pm\,0.18$ & $2.43\,\pm\,0.31$ & $2.47\,\pm\,0.18$ \\
 \hline
\rule{0pt}{3ex} One jet, $15<E_T<25$ GeV & $1.47$ & $\pm\,0.17$ & $ ^{+0.13}_{-0.09}$ & $\pm\,0.09$ & $1.26\,\pm\,0.16$ & $1.18\,\pm\,0.09$ \\ 
\rule{0pt}{3ex} One jet, $25<E_T<45$ GeV & $1.09$ & $\pm\,0.18$ & $ ^{+0.14}_{-0.11}$ & $\pm\,0.06$ & $0.77\,\pm\,0.10$ & $0.79\,\pm\,0.06$ \\ 
\rule{0pt}{3ex} One jet, $E_T>45$ GeV & $0.48$ & $\pm0.15$ & $ ^{+0.19}_{-0.11}$ & $\pm0.03$ & $0.40\pm0.05$ & $0.46\pm0.03$ \\ 
\hline
\rule{0pt}{3ex} Two or more jets & $1.35$ & $\pm\,0.30$ & $ ^{+0.45}_{-0.28}$ & $\pm\,0.08$ & $0.64\,\pm\,0.08$ & $0.61\,\pm\,0.05$ \\
 \hline \hline
\end{tabular}
\caption{Measurements and predictions of $\sigma(p\bar{p} \rightarrow W^{+}W^{-} + \mathrm{jets})$.  Values are given inclusively and differentially as functions of jet multiplicity and jet-transverse energy.}
\label{tab:XS}
\end{table*}

\indent  In summary, the {\it WW} cross section is measured in the dilepton channel both inclusively and differentially in jet multiplicity and $E_T$ using a neutral-net discriminant and binned-maximum-likelihood fit.  
This is the first measurement of the differential cross section for pair production of massive-vector bosons.
The measured cross section, $14.0 \pm 0.6\,(\rm{stat})\, ^{+1.2}_{-1.0}\, (\rm{syst})\, \pm0.8\, (\rm{lumi})$ pb, and the differential cross sections are consistent with both the NLO and fixed-order predictions.   This result indicates the suitability of using either of these theoretical techniques to study processes with multiple gauge bosons and jets. Processes of this type will be used extensively at the LHC to perform searches for non-SM  physics and to investigate the nature of electroweak-symmetry breaking by studying the process of vector-boson scattering.\\
 \indent

We thank the Fermilab staff and the technical staffs of the participating institutions for their vital contributions. This work was supported by the U.S. Department of Energy and National Science Foundation; the Italian Istituto Nazionale di Fisica Nucleare; the Ministry of Education, Culture, Sports, Science and Technology of Japan; the Natural Sciences and Engineering Research Council of Canada; the National Science Council of the Republic of China; the Swiss National Science Foundation; the A.P. Sloan Foundation; the Bundesministerium f\"ur Bildung und Forschung, Germany; the Korean World Class University Program, the National Research Foundation of Korea; the Science and Technology Facilities Council and the Royal Society, UK; the Russian Foundation for Basic Research; the Ministerio de Ciencia e Innovaci\'{o}n, and Programa Consolider-Ingenio 2010, Spain; the Slovak R\&D Agency; the Academy of Finland; the Australian Research Council (ARC); and the EU community Marie Curie Fellowship contract 302103. 

\bibliography{WW_PRD}
\bibliographystyle{apsrev}
\end{document}

%% file: author.tex
\affiliation{Institute of Physics, Academia Sinica, Taipei, Taiwan 11529, Republic of China}
\affiliation{Argonne National Laboratory, Argonne, Illinois 60439, USA}
\affiliation{University of Athens, 157 71 Athens, Greece}
\affiliation{Institut de Fisica d'Altes Energies, ICREA, Universitat Autonoma de Barcelona, E-08193, Bellaterra (Barcelona), Spain}
\affiliation{Baylor University, Waco, Texas 76798, USA}
\affiliation{Istituto Nazionale di Fisica Nucleare Bologna, \ensuremath{^{kk}}University of Bologna, I-40127 Bologna, Italy}
\affiliation{University of California, Davis, Davis, California 95616, USA}
\affiliation{University of California, Los Angeles, Los Angeles, California 90024, USA}
\affiliation{Instituto de Fisica de Cantabria, CSIC-University of Cantabria, 39005 Santander, Spain}
\affiliation{Carnegie Mellon University, Pittsburgh, Pennsylvania 15213, USA}
\affiliation{Enrico Fermi Institute, University of Chicago, Chicago, Illinois 60637, USA}
\affiliation{Comenius University, 842 48 Bratislava, Slovakia; Institute of Experimental Physics, 040 01 Kosice, Slovakia}
\affiliation{Joint Institute for Nuclear Research, RU-141980 Dubna, Russia}
\affiliation{Duke University, Durham, North Carolina 27708, USA}
\affiliation{Fermi National Accelerator Laboratory, Batavia, Illinois 60510, USA}
\affiliation{University of Florida, Gainesville, Florida 32611, USA}
\affiliation{Laboratori Nazionali di Frascati, Istituto Nazionale di Fisica Nucleare, I-00044 Frascati, Italy}
\affiliation{University of Geneva, CH-1211 Geneva 4, Switzerland}
\affiliation{Glasgow University, Glasgow G12 8QQ, United Kingdom}
\affiliation{Harvard University, Cambridge, Massachusetts 02138, USA}
\affiliation{Division of High Energy Physics, Department of Physics, University of Helsinki, FIN-00014, Helsinki, Finland; Helsinki Institute of Physics, FIN-00014, Helsinki, Finland}
\affiliation{University of Illinois, Urbana, Illinois 61801, USA}
\affiliation{The Johns Hopkins University, Baltimore, Maryland 21218, USA}
\affiliation{Institut f\"{u}r Experimentelle Kernphysik, Karlsruhe Institute of Technology, D-76131 Karlsruhe, Germany}
\affiliation{Center for High Energy Physics: Kyungpook National University, Daegu 702-701, Korea; Seoul National University, Seoul 151-742, Korea; Sungkyunkwan University, Suwon 440-746, Korea; Korea Institute of Science and Technology Information, Daejeon 305-806, Korea; Chonnam National University, Gwangju 500-757, Korea; Chonbuk National University, Jeonju 561-756, Korea; Ewha Womans University, Seoul, 120-750, Korea}
\affiliation{Ernest Orlando Lawrence Berkeley National Laboratory, Berkeley, California 94720, USA}
\affiliation{University of Liverpool, Liverpool L69 7ZE, United Kingdom}
\affiliation{University College London, London WC1E 6BT, United Kingdom}
\affiliation{Centro de Investigaciones Energeticas Medioambientales y Tecnologicas, E-28040 Madrid, Spain}
\affiliation{Massachusetts Institute of Technology, Cambridge, Massachusetts 02139, USA}
\affiliation{University of Michigan, Ann Arbor, Michigan 48109, USA}
\affiliation{Michigan State University, East Lansing, Michigan 48824, USA}
\affiliation{Institution for Theoretical and Experimental Physics, ITEP, Moscow 117259, Russia}
\affiliation{University of New Mexico, Albuquerque, New Mexico 87131, USA}
\affiliation{The Ohio State University, Columbus, Ohio 43210, USA}
\affiliation{Okayama University, Okayama 700-8530, Japan}
\affiliation{Osaka City University, Osaka 558-8585, Japan}
\affiliation{University of Oxford, Oxford OX1 3RH, United Kingdom}
\affiliation{Istituto Nazionale di Fisica Nucleare, Sezione di Padova, \ensuremath{^{ll}}University of Padova, I-35131 Padova, Italy}
\affiliation{University of Pennsylvania, Philadelphia, Pennsylvania 19104, USA}
\affiliation{Istituto Nazionale di Fisica Nucleare Pisa, \ensuremath{^{mm}}University of Pisa, \ensuremath{^{nn}}University of Siena, \ensuremath{^{oo}}Scuola Normale Superiore, I-56127 Pisa, Italy, \ensuremath{^{pp}}INFN Pavia, I-27100 Pavia, Italy, \ensuremath{^{qq}}University of Pavia, I-27100 Pavia, Italy}
\affiliation{University of Pittsburgh, Pittsburgh, Pennsylvania 15260, USA}
\affiliation{Purdue University, West Lafayette, Indiana 47907, USA}
\affiliation{University of Rochester, Rochester, New York 14627, USA}
\affiliation{The Rockefeller University, New York, New York 10065, USA}
\affiliation{Istituto Nazionale di Fisica Nucleare, Sezione di Roma 1, \ensuremath{^{rr}}Sapienza Universit\`{a} di Roma, I-00185 Roma, Italy}
\affiliation{Mitchell Institute for Fundamental Physics and Astronomy, Texas A\&M University, College Station, Texas 77843, USA}
\affiliation{Istituto Nazionale di Fisica Nucleare Trieste, \ensuremath{^{ss}}Gruppo Collegato di Udine, \ensuremath{^{tt}}University of Udine, I-33100 Udine, Italy, \ensuremath{^{uu}}University of Trieste, I-34127 Trieste, Italy}
\affiliation{University of Tsukuba, Tsukuba, Ibaraki 305, Japan}
\affiliation{Tufts University, Medford, Massachusetts 02155, USA}
\affiliation{University of Virginia, Charlottesville, Virginia 22906, USA}
\affiliation{Waseda University, Tokyo 169, Japan}
\affiliation{Wayne State University, Detroit, Michigan 48201, USA}
\affiliation{University of Wisconsin, Madison, Wisconsin 53706, USA}
\affiliation{Yale University, New Haven, Connecticut 06520, USA}

\author{T.~Aaltonen}
\affiliation{Division of High Energy Physics, Department of Physics, University of Helsinki, FIN-00014, Helsinki, Finland; Helsinki Institute of Physics, FIN-00014, Helsinki, Finland}
\author{S.~Amerio\ensuremath{^{ll}}}
\affiliation{Istituto Nazionale di Fisica Nucleare, Sezione di Padova, \ensuremath{^{ll}}University of Padova, I-35131 Padova, Italy}
\author{D.~Amidei}
\affiliation{University of Michigan, Ann Arbor, Michigan 48109, USA}
\author{A.~Anastassov\ensuremath{^{w}}}
\affiliation{Fermi National Accelerator Laboratory, Batavia, Illinois 60510, USA}
\author{A.~Annovi}
\affiliation{Laboratori Nazionali di Frascati, Istituto Nazionale di Fisica Nucleare, I-00044 Frascati, Italy}
\author{J.~Antos}
\affiliation{Comenius University, 842 48 Bratislava, Slovakia; Institute of Experimental Physics, 040 01 Kosice, Slovakia}
\author{G.~Apollinari}
\affiliation{Fermi National Accelerator Laboratory, Batavia, Illinois 60510, USA}
\author{J.A.~Appel}
\affiliation{Fermi National Accelerator Laboratory, Batavia, Illinois 60510, USA}
\author{T.~Arisawa}
\affiliation{Waseda University, Tokyo 169, Japan}
\author{A.~Artikov}
\affiliation{Joint Institute for Nuclear Research, RU-141980 Dubna, Russia}
% ...

\collaboration{CDF Collaboration}
\altaffiliation[With visitors from]{
\ensuremath{^{a}}University of British Columbia, Vancouver, BC V6T 1Z1, Canada,
\ensuremath{^{b}}Istituto Nazionale di Fisica Nucleare, Sezione di Cagliari, 09042 Monserrato (Cagliari), Italy,
\ensuremath{^{c}}University of California Irvine, Irvine, CA 92697, USA,
\ensuremath{^{d}}Institute of Physics, Academy of Sciences of the Czech Republic, 182~21, Czech Republic,
\ensuremath{^{e}}CERN, CH-1211 Geneva, Switzerland,
\ensuremath{^{f}}Cornell University, Ithaca, NY 14853, USA,
\ensuremath{^{g}}University of Cyprus, Nicosia CY-1678, Cyprus,
\ensuremath{^{h}}Office of Science, U.S. Department of Energy, Washington, DC 20585, USA,
\ensuremath{^{i}}University College Dublin, Dublin 4, Ireland,
\ensuremath{^{j}}ETH, 8092 Z\"{u}rich, Switzerland,
\ensuremath{^{k}}University of Fukui, Fukui City, Fukui Prefecture, Japan 910-0017,
\ensuremath{^{l}}Universidad Iberoamericana, Lomas de Santa Fe, M\'{e}xico, C.P. 01219, Distrito Federal,
\ensuremath{^{m}}University of Iowa, Iowa City, IA 52242, USA,
\ensuremath{^{n}}Kinki University, Higashi-Osaka City, Japan 577-8502,
\ensuremath{^{o}}Kansas State University, Manhattan, KS 66506, USA,
\ensuremath{^{p}}Brookhaven National Laboratory, Upton, NY 11973, USA,
\ensuremath{^{q}}Istituto Nazionale di Fisica Nucleare, Sezione di Lecce, Via Arnesano, I-73100 Lecce, Italy,
\ensuremath{^{r}}Queen Mary, University of London, London, E1 4NS, United Kingdom,
\ensuremath{^{s}}University of Melbourne, Victoria 3010, Australia,
\ensuremath{^{t}}Muons, Inc., Batavia, IL 60510, USA,
\ensuremath{^{u}}Nagasaki Institute of Applied Science, Nagasaki 851-0193, Japan,
\ensuremath{^{v}}National Research Nuclear University, Moscow 115409, Russia,
\ensuremath{^{w}}Northwestern University, Evanston, IL 60208, USA,
\ensuremath{^{x}}University of Notre Dame, Notre Dame, IN 46556, USA,
\ensuremath{^{y}}Universidad de Oviedo, E-33007 Oviedo, Spain,
\ensuremath{^{z}}CNRS-IN2P3, Paris, F-75205 France,
\ensuremath{^{aa}}Universidad Tecnica Federico Santa Maria, 110v Valparaiso, Chile,
\ensuremath{^{bb}}Sejong University, Seoul, South Korea,
\ensuremath{^{cc}}The University of Jordan, Amman 11942, Jordan,
\ensuremath{^{dd}}Universite catholique de Louvain, 1348 Louvain-La-Neuve, Belgium,
\ensuremath{^{ee}}University of Z\"{u}rich, 8006 Z\"{u}rich, Switzerland,
\ensuremath{^{ff}}Massachusetts General Hospital, Boston, MA 02114 USA,
\ensuremath{^{gg}}Harvard Medical School, Boston, MA 02114 USA,
\ensuremath{^{hh}}Hampton University, Hampton, VA 23668, USA,
\ensuremath{^{ii}}Los Alamos National Laboratory, Los Alamos, NM 87544, USA,
\ensuremath{^{jj}}Universit\`{a} degli Studi di Napoli Federico I, I-80138 Napoli, Italy
}
\noaffiliation